\documentclass[pra,showpacs,floatfix,12pt]{revtex4}
\usepackage{graphicx}

\begin{document}

\title{Colloid--Polymer Mixtures Confined Between Asymmetric Walls:
Simulation Evidence for an Interface Localization Transition}

\author{A. De Virgiliis$^1$, R.L.C. Vink$^{1,2}$, J. Horbach$^1$, and K. Binder$^1$}

\affiliation{$^1$Institut f\"{u}r Physik, Johannes--Gutenberg--Universit\"{a}t,
D-55099 Mainz, Staudinger Weg 7, Germany\\
$^2$Institut f\"ur Theoretische Physik II, Heinrich--Heine--Universit\"at
D\"usseldorf, Universit\"atsstr.~1, 40225 D\"usseldorf, Germany}

\begin{abstract}
Phase separation of colloid--polymer mixtures, described by the
Asakura--Oosawa (AO) model, confined in a thin slit pore is studied by
grand--canonical Monte Carlo simulation. While one wall is a hard wall
for both particles, at the other wall there acts an additional repulsive
potential on the colloids only. Varying the strength of this potential,
a crossover from capillary condensation to an interface localization
transition is found. The latter occurs under conditions where in the
bulk almost complete phase separation has occurred.
\end{abstract}

\maketitle

\section{Introduction and motivation}
The current paradigm to create nanoscopic devices has enhanced the
interest in the changes of phase behavior and structure when fluids are
confined to linear dimensions that are no longer very large compared to
the particle sizes. E.g., for a fluid (or fluid mixture) in a nanoscopic
slit pore an interesting interplay of finite size effects, wetting or
drying, and capillary condensation or evaporation--like behavior can
be expected \cite{1,2,3,4,5,7,8,9,10,11,12,13,14,15,16,17,18,19}. A
particularly interesting theoretical prediction \cite{3,4,5,7,8} concerns
systems that can undergo phase separation (a fluid separating into vapor
and liquid, or a binary mixture with a miscibility gap) confined between
``competing walls". By competing walls we mean that one wall favors one
of the coexisting phases and the other wall favors the other one (the
generic model \cite{3,4,5,7,8} is a thin ferromagnetic Ising film with
oppositely oriented surface fields at both surfaces). When one brings the
system in a state where phase separation occurs in the bulk, the thin film
also exhibits phase separation, with an interface parallel to the walls
in the center of the film \cite{3,4,5,7,8,10,11,12,13,16,17,20}. However,
sufficiently far off from bulk criticality then a transition occurs where
the interface gets localized at the wall(s). This transition can be of
first order or of second order \cite{3,4,5,7,8} and in the latter case it
was argued to belong to the universality class of the two--dimensional
($2d$) Ising model. However, while interfaces parallel to the walls of
thin films could be observed experimentally in thin films of polymer
mixtures \cite{20}, no experimental observation of this novel type of
transition has as yet been reported. Indeed, the theoretical models have
always invoked perfect symmetry between the two coexisting phases in
the bulk (spin reversal symmetry of the Ising magnet \cite{3,4,5,7,8},
or strictly symmetric polymer blends \cite{10,11,12,13,16}, which are
hard to realize in nature).

In the present work we show that this interface localization should
also be observable in colloid--polymer mixtures, which are known to be
model systems for the experimental study of phase separation \cite{21}
and interfacial fluctuations \cite{22,23}.  The large size of the
colloid particles (of the order of 1\,$\mu$m) should allow to prepare
slit pores which are of the order of 10 -- 100 particle diameters wide,
with strongly repulsive walls which are essentially perfectly flat on
this mesoscopic scale, modeling thus a ``hard wall'' boundary. However,
coating a wall with a layer of long endgrafted flexible polymers of the
same chemical type as used in the colloid--polymer mixture, one could
create an additional repulsive interaction to the colloids to cancel
(partially or fully) the depletion attraction \cite{15,24} of the colloids
to the hard walls. Varying the grafting density in the resulting ``polymer
brush'' \cite{25} and/or the molecular weight of the grafted chains the
range and strength of this short range repulsion between colloids and
the wall could be fine--tuned, within some limits \cite{26}.

In the following we shall present model calculations of a simple model
for colloid--polymer mixtures, the Asakura--Oosawa (AO) model \cite{27},
confined between two inequivalent walls: a hard wall on one side, and
a hard wall plus short range repulsion acting only on the colloids on
the other side. Varying the strength of this repulsion we demonstrate
a crossover from capillary condensation--like behavior to interface
localization transitions.

\section{Model and simulation technique}
The colloids are modeled as hard spheres with diameter $\sigma_{\rm
c}$, the polymers are (soft) spheres with diameter $\sigma_{\rm
p}$. The polymers may not overlap with colloids, but there is no
interaction between the polymers. Recall that flexible polymers in
solution form random coils with a rather large gyration radius which
may interpenetrate at very low energy cost.  We use a size ratio
$q=\sigma_{\rm p}/\sigma_{\rm c}=0.8$, since the phase diagram of this
model in the bulk has been carefully studied previously \cite{28,29}. We
choose $\sigma_{\rm c}=1.0$ as our unit of length.  When there are
$N_{\rm p}$ polymers and $N_{\rm c}$ colloids in the considered
volume $V$, the polymer and colloid packing fractions are defined as
$\eta_{\rm p}=\pi \sigma_{\rm p}^3 N_{\rm p}/(6V)$ and $\eta_{\rm c}=
\pi \sigma_{\rm c}^3 N_{\rm c}/(6V)$, respectively. The volume is given
by $V=L \times L \times D$.  Here, $D$ is the width of the slit pore
in $z$--direction. Periodic boundary conditions are used in $x$-- and
$y$--directions parallel to the walls. The $L \times L$ wall at $z=0$
simply is a hard wall, while the wall at $z=D$ is described by an
additional repulsive potential.  It has the following form (we absorb
a factor $k_BT$ here, $k_B$=Boltzmann's constant, $T$=temperature):
\begin{eqnarray} \label{eq1}
u_{\rm cw}(z)=  \left \{
\begin{array}{lll}
\infty \quad {\rm for}\quad   z\leq 0\\
\varepsilon \quad  \,\,{\rm for} \quad  0 <z \leq \sigma_{\rm c}/2\\
0 \quad  \,\, {\rm otherwise.}
\end{array} \right .
\end{eqnarray}
The strength $\varepsilon$ of this wall--colloid repulsion is
varied between $0$ and $2.5$, while for polymers both walls are
purely repulsive. Slit widths $D=5$ and $D=10$ were studied, while
$L$ was varied from $L=10$ to $L=30$, in the framework of a finite
size scaling \cite{30} analysis.

\begin{figure}
\centering
\includegraphics*[width=.4\textwidth]{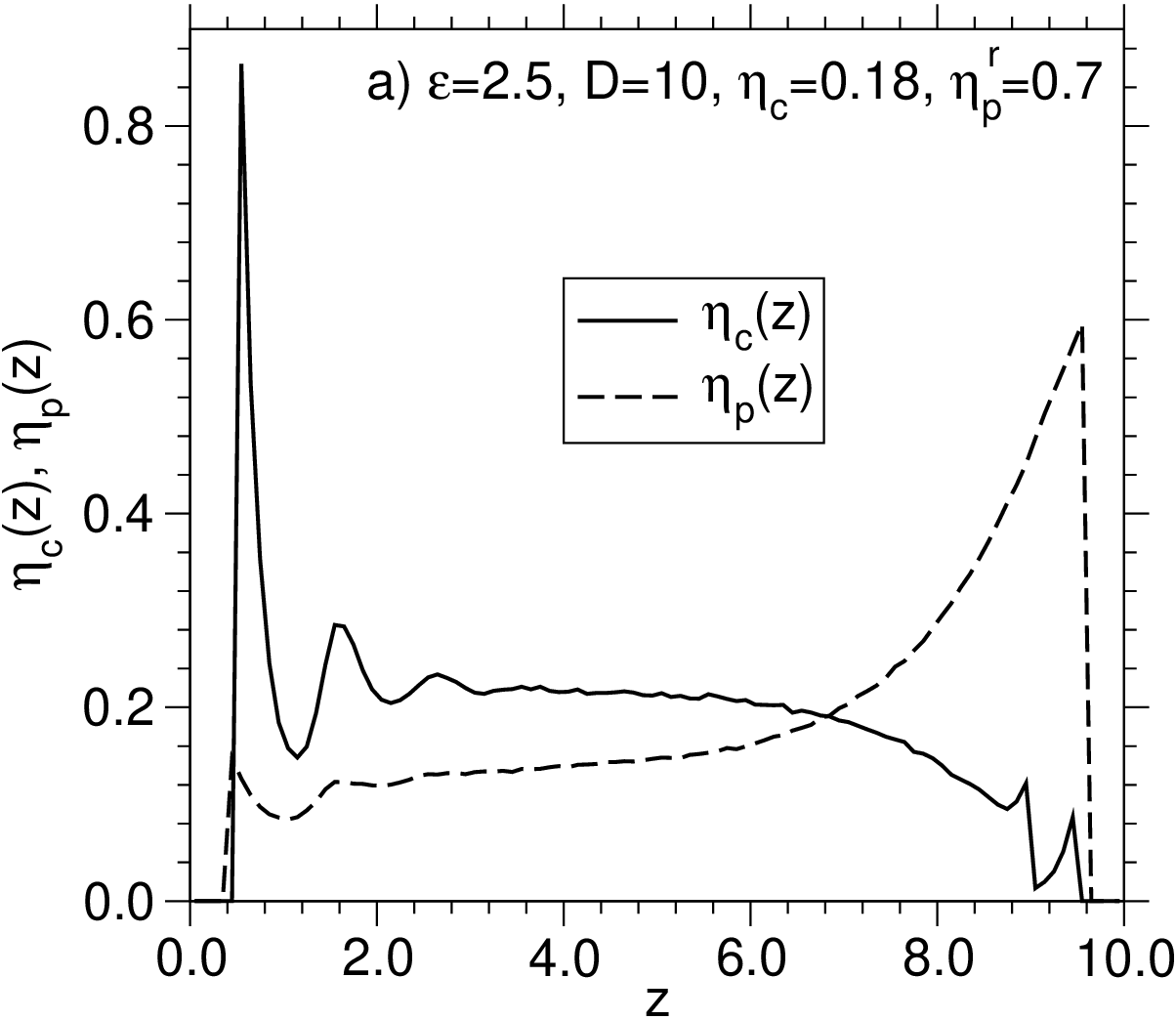}
\includegraphics*[width=.4\textwidth]{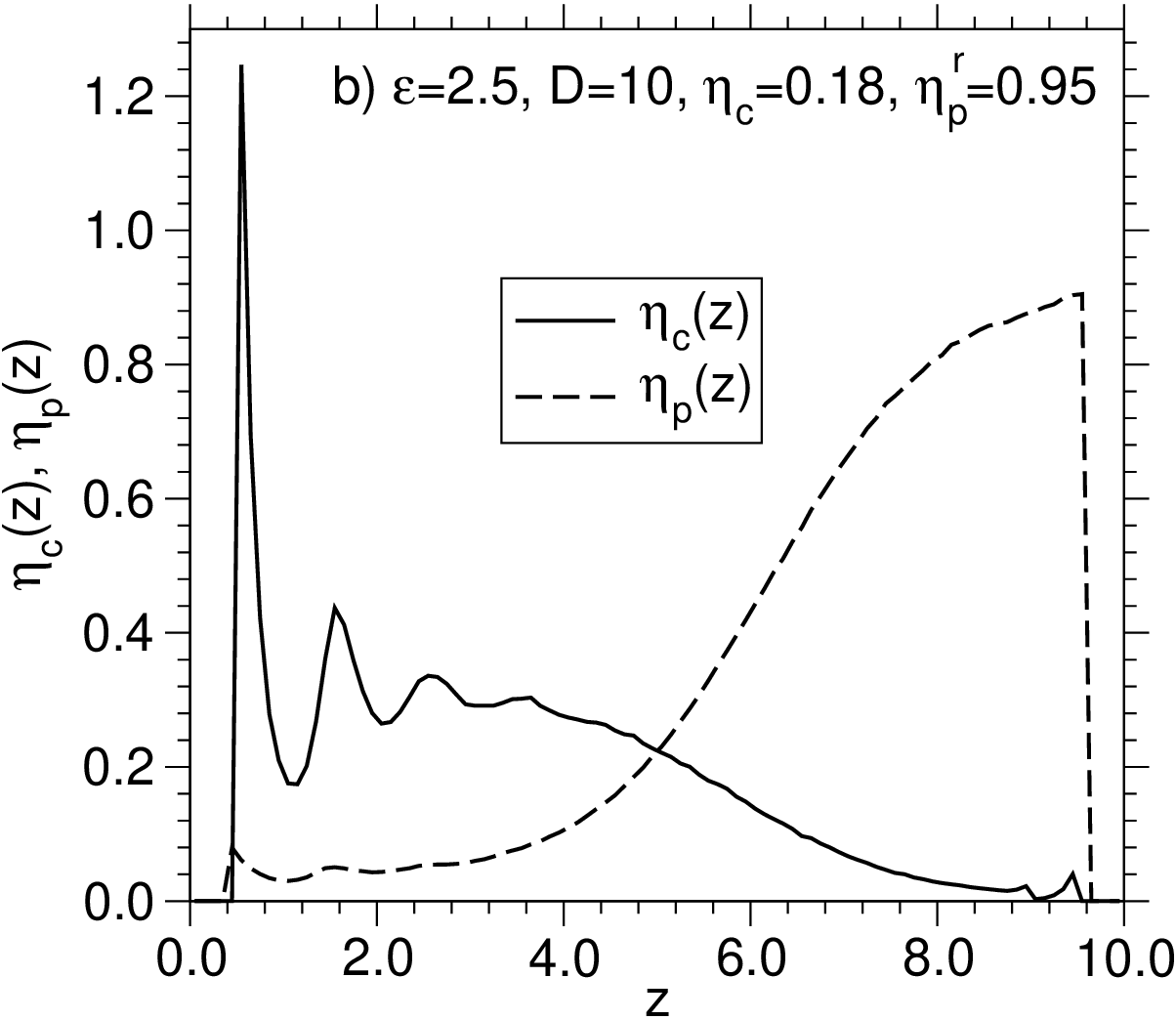}
\includegraphics*[width=.4\textwidth]{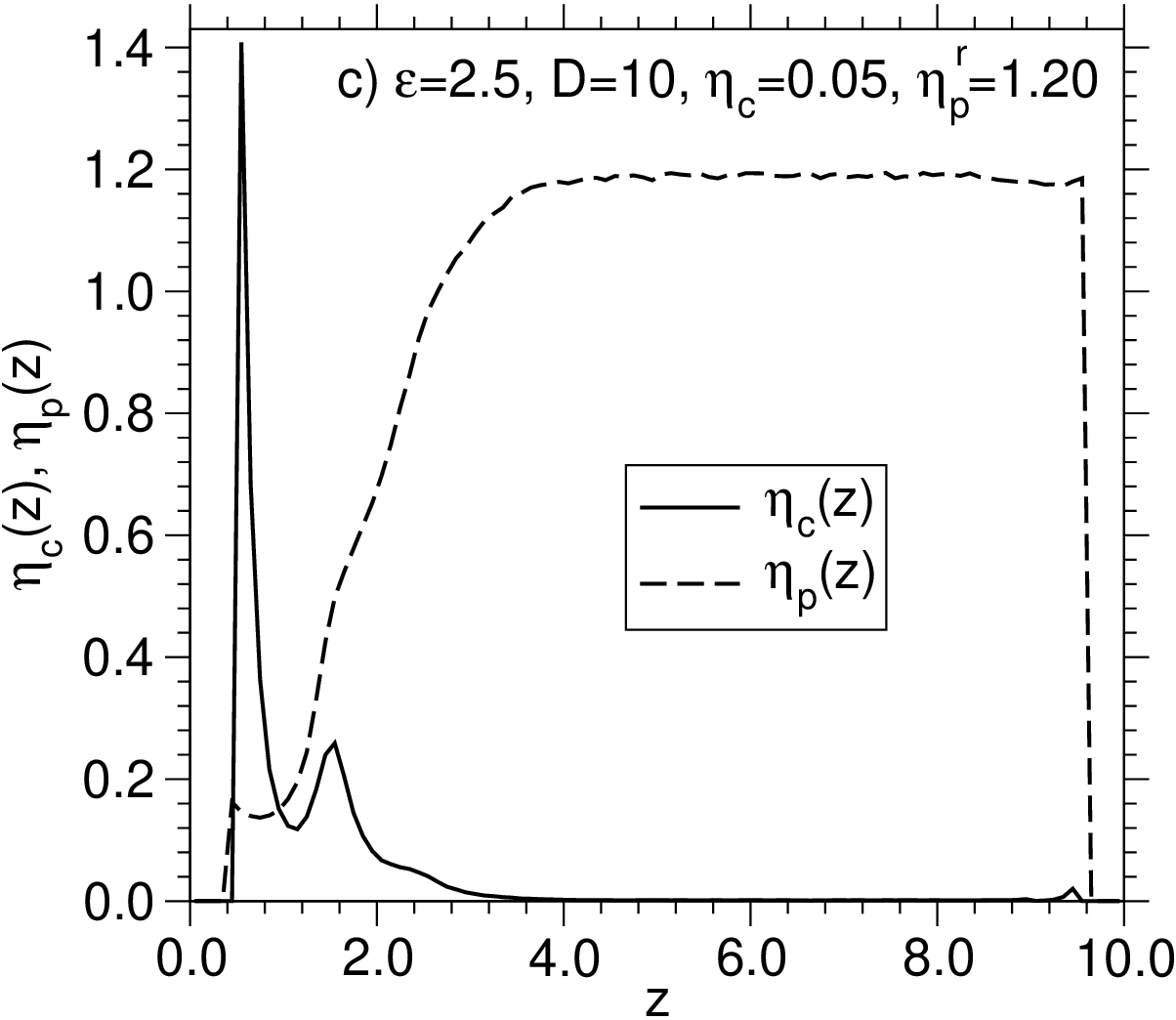}
\includegraphics*[width=.4\textwidth]{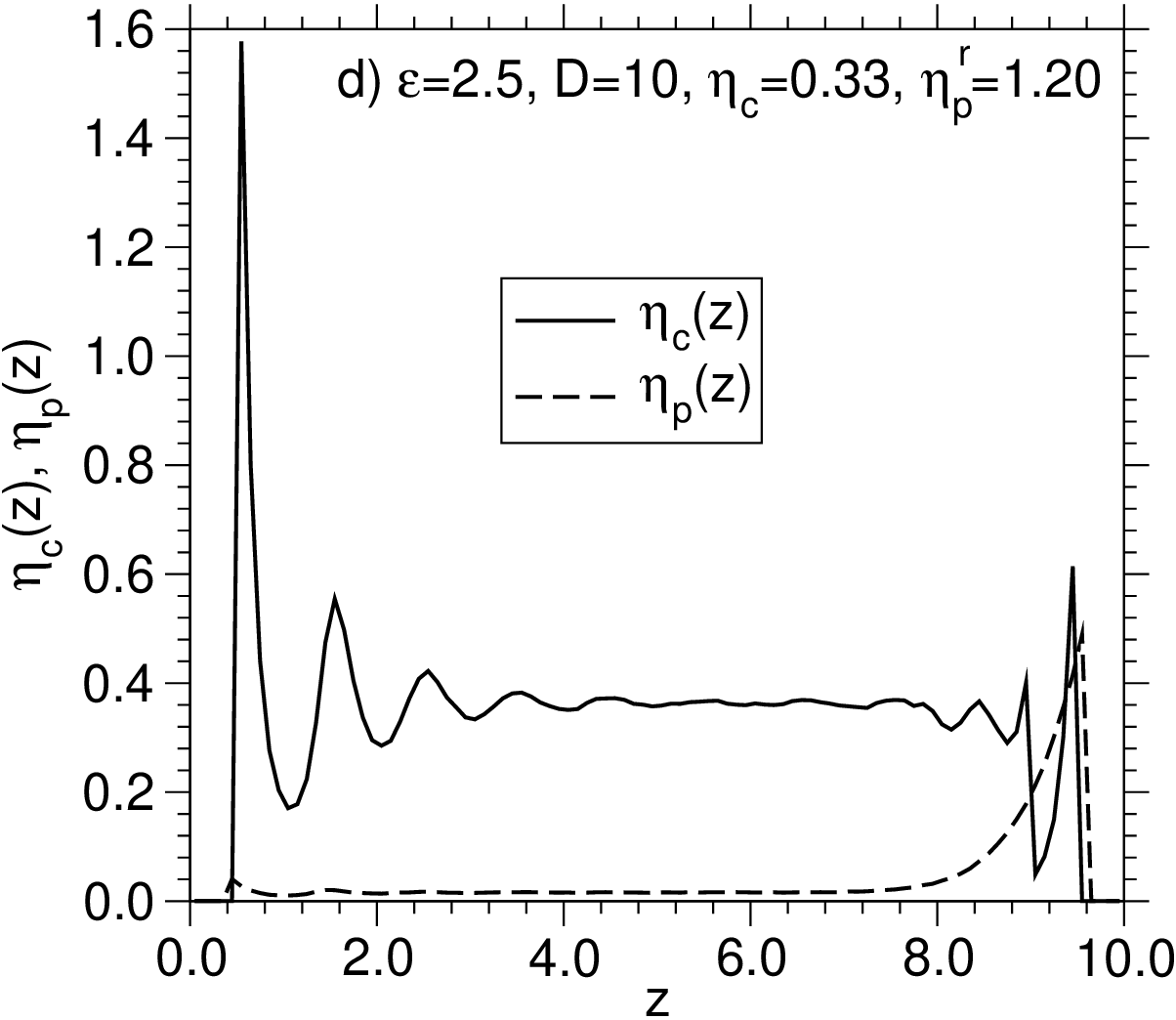}
\caption{Colloid concentration profiles $\eta_{\rm c}(z)$ and polymer
concentration profiles $\eta_{\rm p}(z)$ as function of $z$ for a
thin film with asymmetric walls (hard wall at $z=0$, while for the
other wall at $z=D=10$ the potential of Eq.~(\ref{eq1}) acts, with
$\varepsilon=2.5$). Profiles were obtained at $\eta_{\rm c}=0.18$,
$\eta_{\rm p}^{\rm r}=0.70$ (a), $\eta_{\rm c}=0.18$, $\eta^{\rm r}_{\rm
p}=0.95$ (b), $\eta_{\rm c}=0.05$, $\eta_{\rm p}^{\rm r}=1.20$ (c), and
$\eta_{\rm c}=0.33$, $\eta_{\rm p}^{\rm r} =1.20$ (d). For profiles (c)
and (d), the choices $\eta_{\rm c}=0.05, 0.33$ roughly correspond to the
two branches of the coexistence curve, see Fig.~\ref{fig2}. \label{fig1}}
\end{figure}
\begin{figure}
\centering
\includegraphics*[width=.45\textwidth]{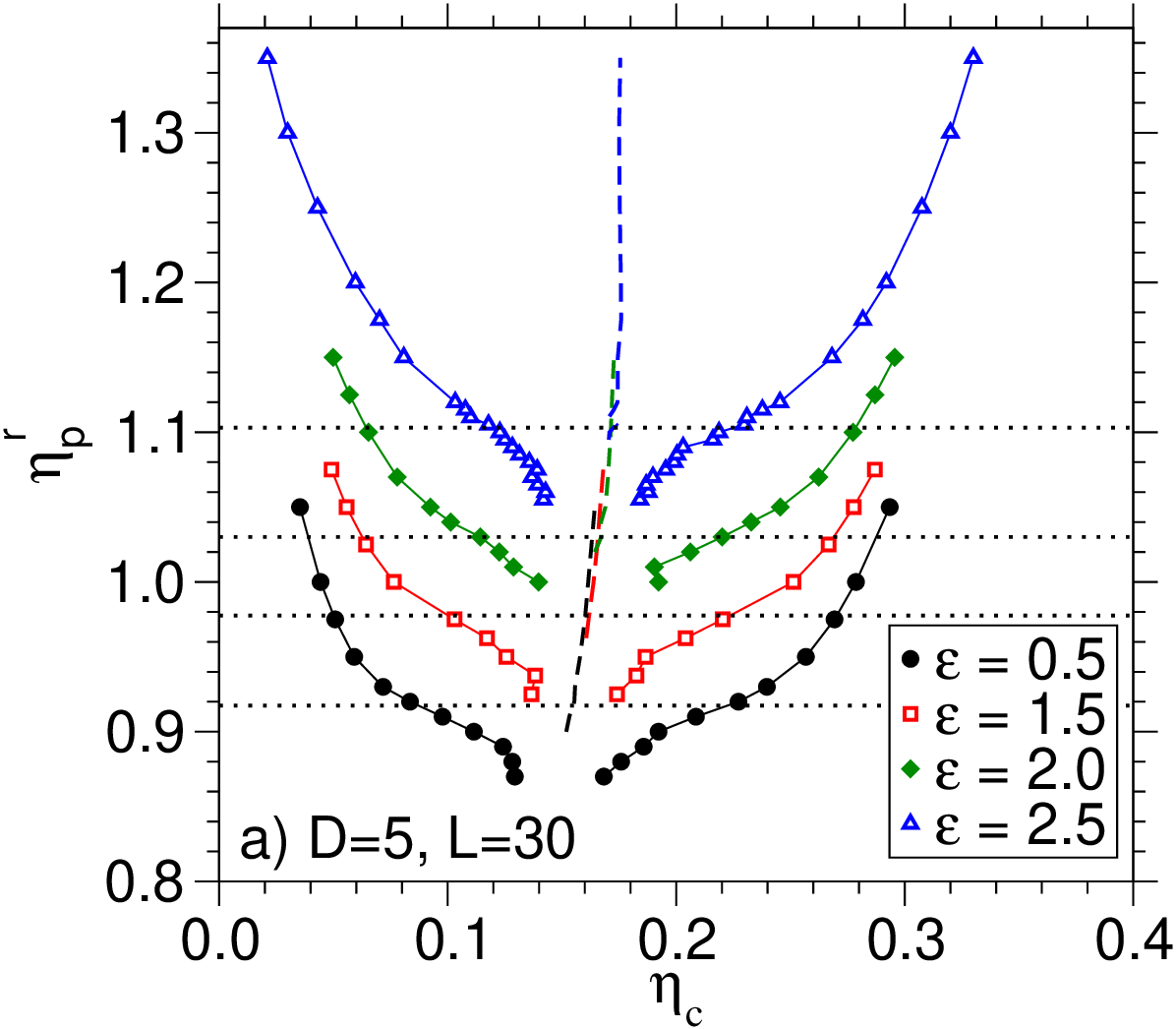}
\includegraphics*[width=.45\textwidth]{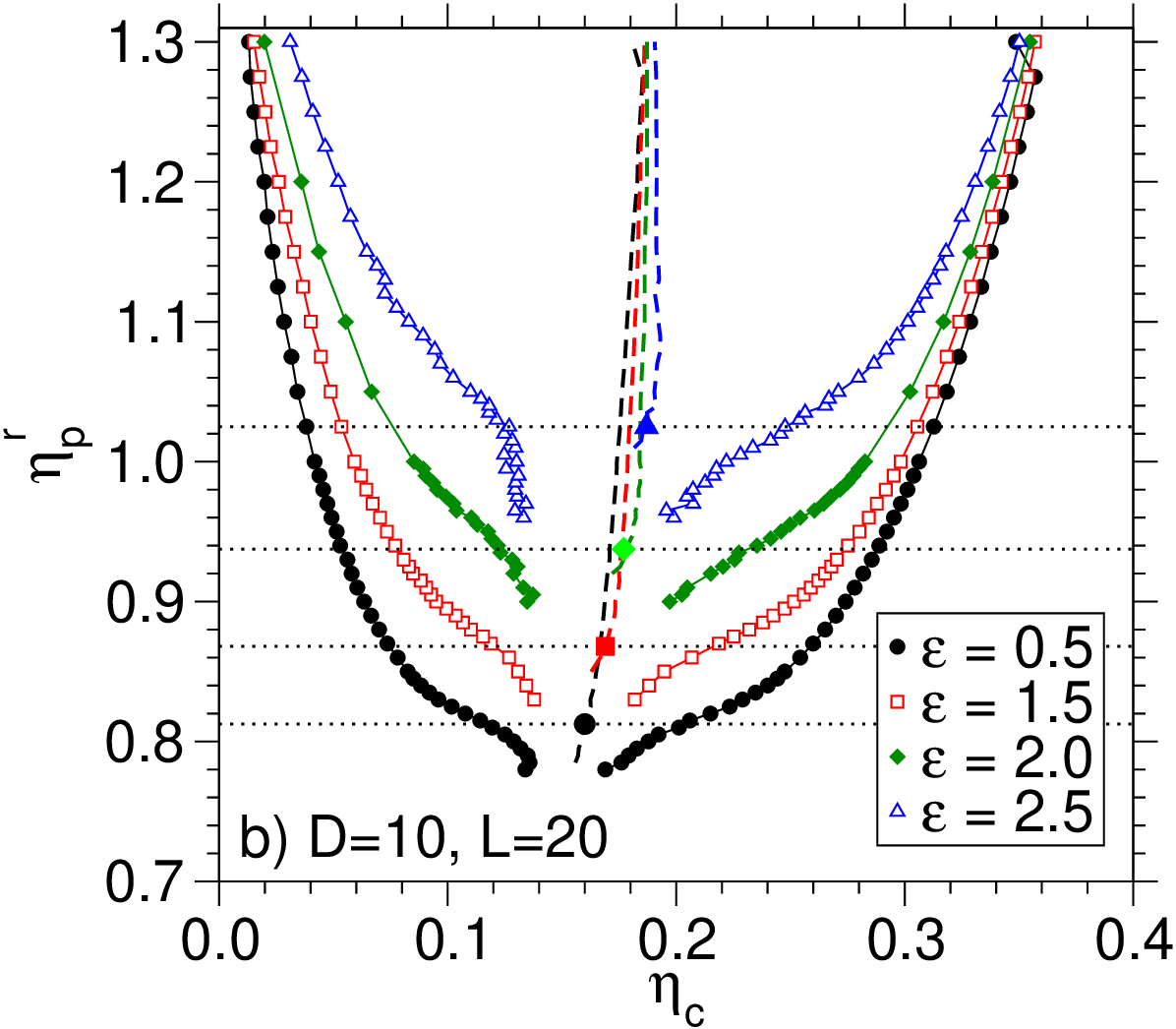}
\caption{Coexistence curves for $D=5$, $L=30$ (a) and $D=10$, $L=20$ (b),
showing four choices of $\varepsilon$ in each case. The data points with
open symbols show the vapor--like $(\eta^v_{\rm c})$ and liquid--like
$(\eta^\ell_{\rm c}$) branches of the coexistence curve, as extracted
from the peaks of $P_L$. The broken curves show the coexistence diameter
$\delta=(\eta^v_{\rm c} + n_{\rm c}^\ell)/2$, on which the critical points
are marked with full symbols. The critical values $\eta^{\rm r}_{\rm p,
crit}$ of $\eta_{\rm p}^{\rm r}$ are highlighted by dotted horizontal
lines. \label{fig2}}
\end{figure}
The simulation is carried out in the grand--canonical ensemble,
with the chemical potentials $\mu_{\rm p}$, $\mu_{\rm c}$ of polymers
and colloids (or fugacities $z_{\rm p}$, $z_{\rm c}$) as independent
variables.  Following common practice, we use the so--called ``polymer
reservoir packing fraction'' $\eta^{\rm r}_{\rm p} \equiv \pi z_{\rm p}
\sigma^3_{\rm p}/6$ rather than $z_{\rm p}$ as the temperature--like
variable.  As in the study of bulk critical behavior \cite{28,29} and
``capillary condensation'' \cite{18,19} of the colloids on hard walls
we use a grand--canonical cluster move \cite{27} together with a very
efficient reweighting scheme, successive umbrella sampling \cite{31}, to
obtain the distribution function $P_L (\eta_{\rm c}|n^{\rm r}_{\rm p},\,
z_{\rm c})$. This distribution function is defined as the probability
to observe the system with colloid packing fraction $\eta_{\rm c}$ at
``inverse temperature'' $\eta^{\rm r}_{\rm p}$ and colloid fugacity
$\eta_{\rm c}$. For states far away from phase coexistence, $P_L$ is a
single--peaked function, while near phase coexistence a double--peak
structure develops \cite{18,19,28,29,30,31}. The precise location of
the value of $z_{\rm c}$ at which two--phase coexistence occurs is
given by the equal weight rule \cite{32}. The positions of the two
peaks of $P_L$ then yield (preliminary) estimates for the two branches
of the coexistence curve, the ``liquid'' branch $\eta_{\rm c}^{\ell}$ and
the ``vapor'' branch $\eta_{\rm c}^{v}$.  However, near criticality these
estimates are affected by finite size effects \cite{28,29,30}.  To deal
with the latter, we study reduced moments of $P_L$ at phase coexistence,
defining an analogue of the order parameter of the Ising model,
\begin{equation} \label{eq2}
m=\eta_{\rm c}-\langle \eta_{\rm c}\rangle, \quad 
\langle \eta_{\rm c} \rangle
=\int\limits_0^\infty \, \eta_{\rm c} 
P_L (\eta_{\rm c}|\eta_{\rm p}^{\rm r}, 
\, z_{\rm c}) d \eta_{\rm c} \; ,
\end{equation}
and higher order moments, $\langle m^p \rangle = \int m^p \,\,
P_L (\eta_{\rm c}| \eta_{\rm p}^{\rm r}, \, z_{\rm c})d\eta_{\rm
c}$. Following the behavior of ratios such as $U_4 = \langle m^2
\rangle^2/\langle m^4 \rangle$ along the path in the $(z_{\rm c},
\eta_{\rm p}^{\rm r})$ phase along which phase coexistence occurs for
several choices of $L$ one can estimate the critical point (``cumulant
intersection method'' \cite{30}).

\section{Simulation results}
\begin{figure}
\centering
\includegraphics*[width=.45\textwidth]{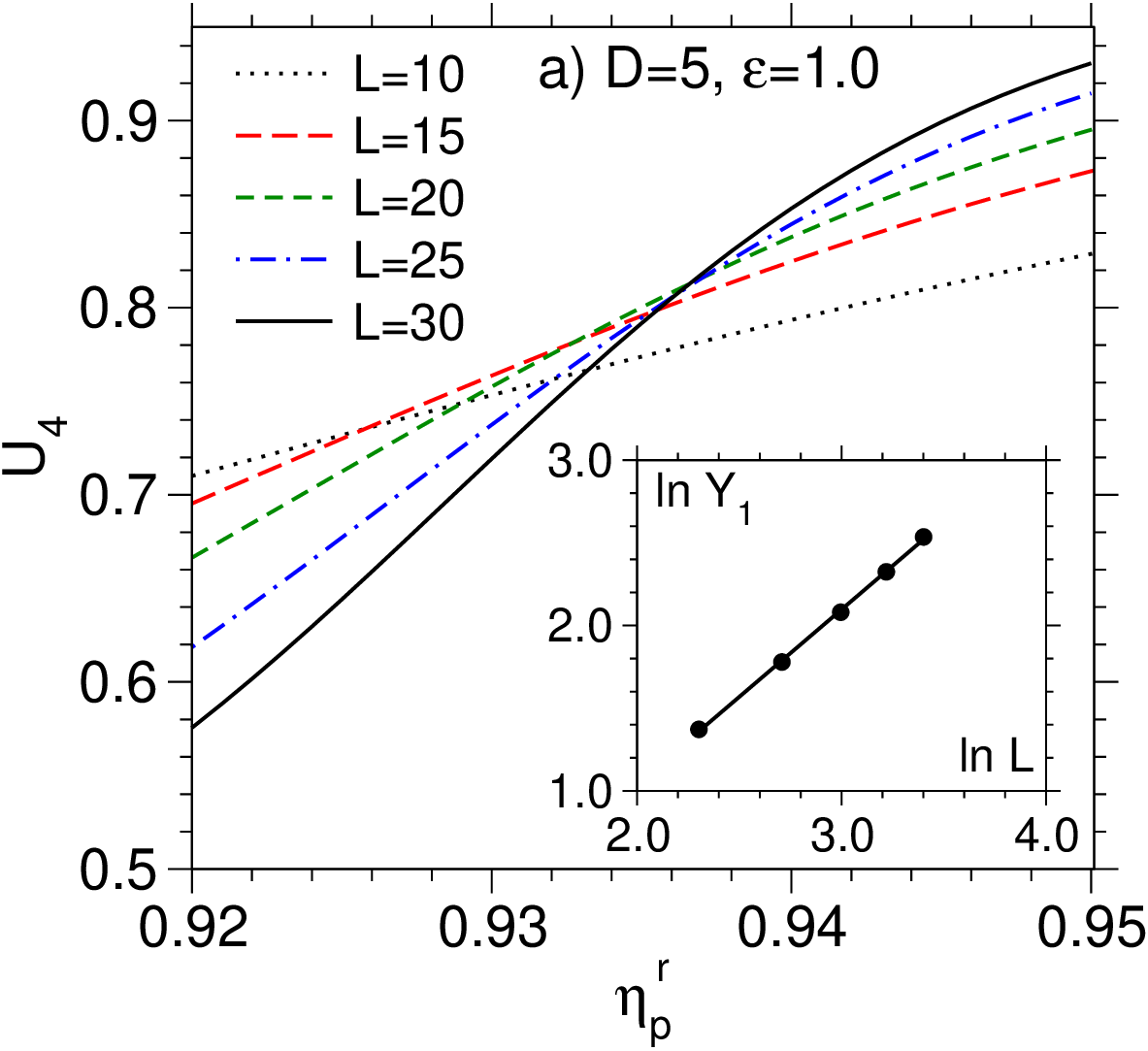}
\includegraphics*[width=.45\textwidth]{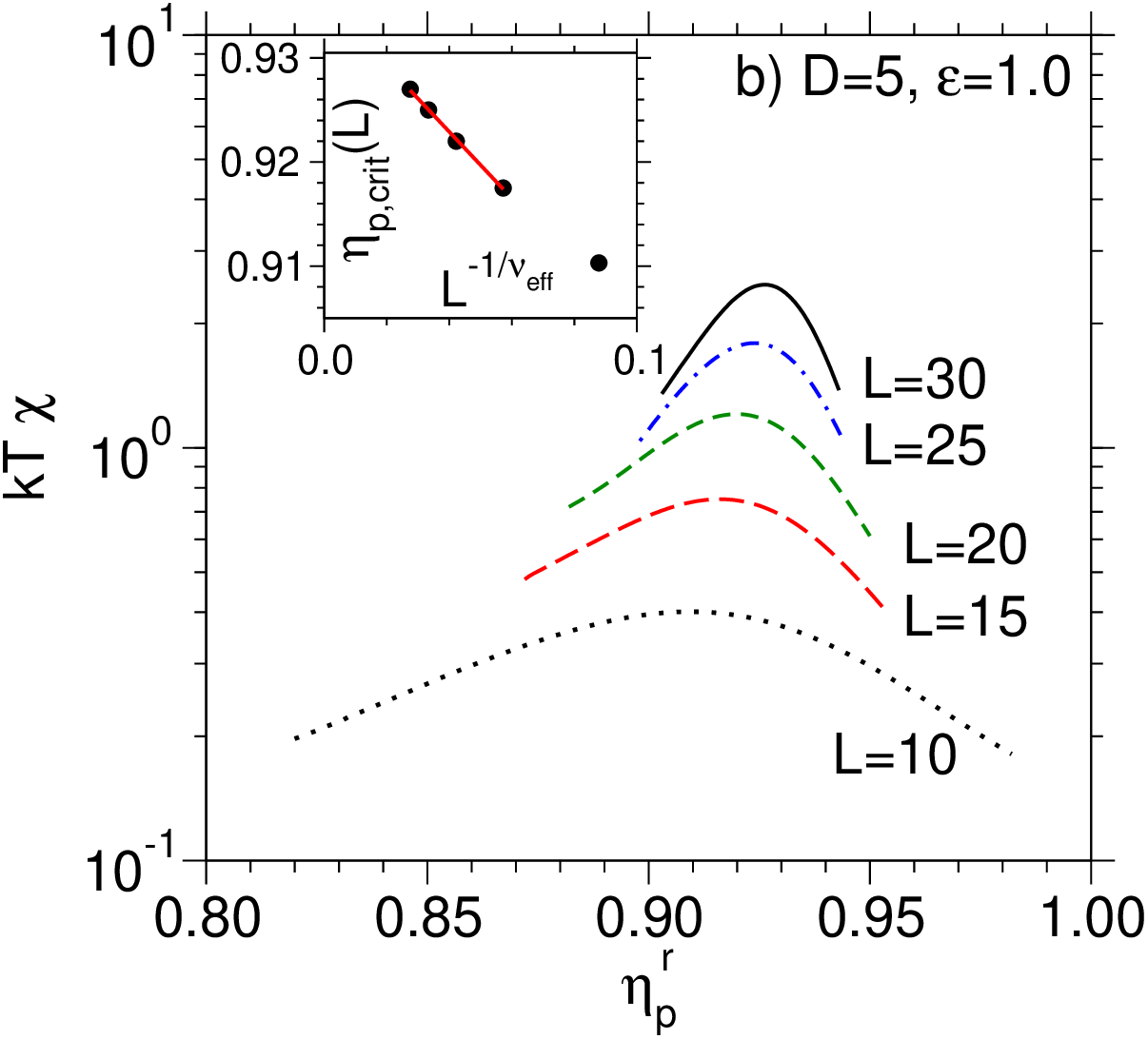}
\caption{Fourth--order cumulant $U_4$ (a) and susceptibility (b)
$k_BT\chi=L^2D (\langle m^2 \rangle - \langle |m| \rangle^2 )$
plotted vs. $\eta^{\rm r}_{\rm p}$, for a slit of thickness $D=5$
and $\varepsilon=1.0$, for several values of $L$ as indicated in the
figure. Insert in a) shows the slope $Y_1$ of the cumulants, evaluated
at $\eta^{\rm r}_{\rm p, crit}=0.937 \pm 0.003$, on log--log scales to
extract the exponent $\nu$ (note \cite{30} $Y_1 \propto L^{1/\nu})$,
yielding $1/\nu \approx 1.056$. The inset in b) shows the extrapolation
of the peak positions, from which $\eta^{\rm r}_{\rm p, crit}=0.935\pm
0.005$ results, consistent with (a). \label{fig3}}
\end{figure}
Figure~\ref{fig1} shows typical concentration profiles for a slit of width
$D=10$ at $\eta_{\rm c}=0.18$ and $\varepsilon=2.5$ and three choices
of the polymer reservoir packing fraction: $\eta_{\rm p}^{\rm r}=0.7$
(in the one phase region of the bulk), $\eta_{\rm p}^{\rm r}=0.95$ (in
the two phase region of the bulk) and at $\eta^{\rm r}_{\rm p}=1.20$
(as we shall see, this is in the two phase region of the thin film:
therefore two profiles are shown here, corresponding to the two
coexisting phases). For $\eta^{\rm r}_{\rm p}=0.7$ one can see that
there is an enhancement of the colloid concentration on the hard wall, as
expected from the depletion attraction already noted in previous studies
\cite{14,15,18,19}.  On the repulsive wall, the colloid concentration
is somewhat depressed (and the polymer concentration slightly enhanced),
but in the center of the film both concentrations are roughly constant,
as expected for bulk--like behavior.

\begin{figure}
\centering
\includegraphics*[width=.5\textwidth]{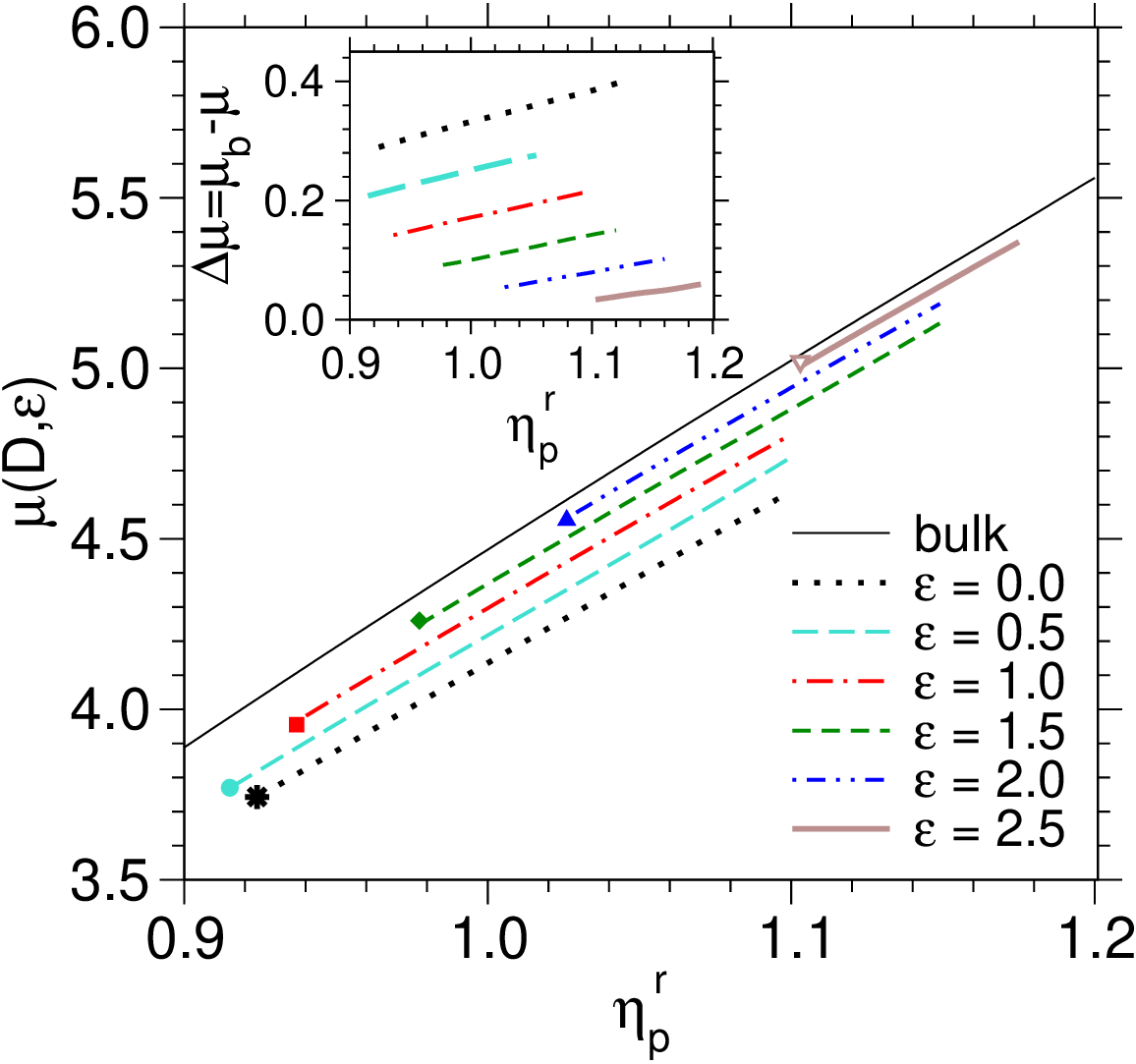}
\caption{Phase boundary $\mu_{\rm coex}$ $(D, \varepsilon)$ plotted
vs.~$\eta_{\rm p}^{\rm r}$ for several choices of $\varepsilon$. For
comparison, also the symmetric slit, $\varepsilon=0$ \cite{18}
and the bulk phase boundary $\mu^{\rm bulk}_{\rm coex}$ \cite{28}
are included. Full symbols mark the locations of the critical points.
Inset shows $\Delta\mu=\mu^{\rm bulk}_{\rm coex}-\mu_{\rm coex}$ $(D,
\varepsilon)$ versus $\eta^{\rm r}_{\rm p}$. \label{fig4}}
\end{figure}
\begin{figure}
\centering
\includegraphics*[width=.5\textwidth]{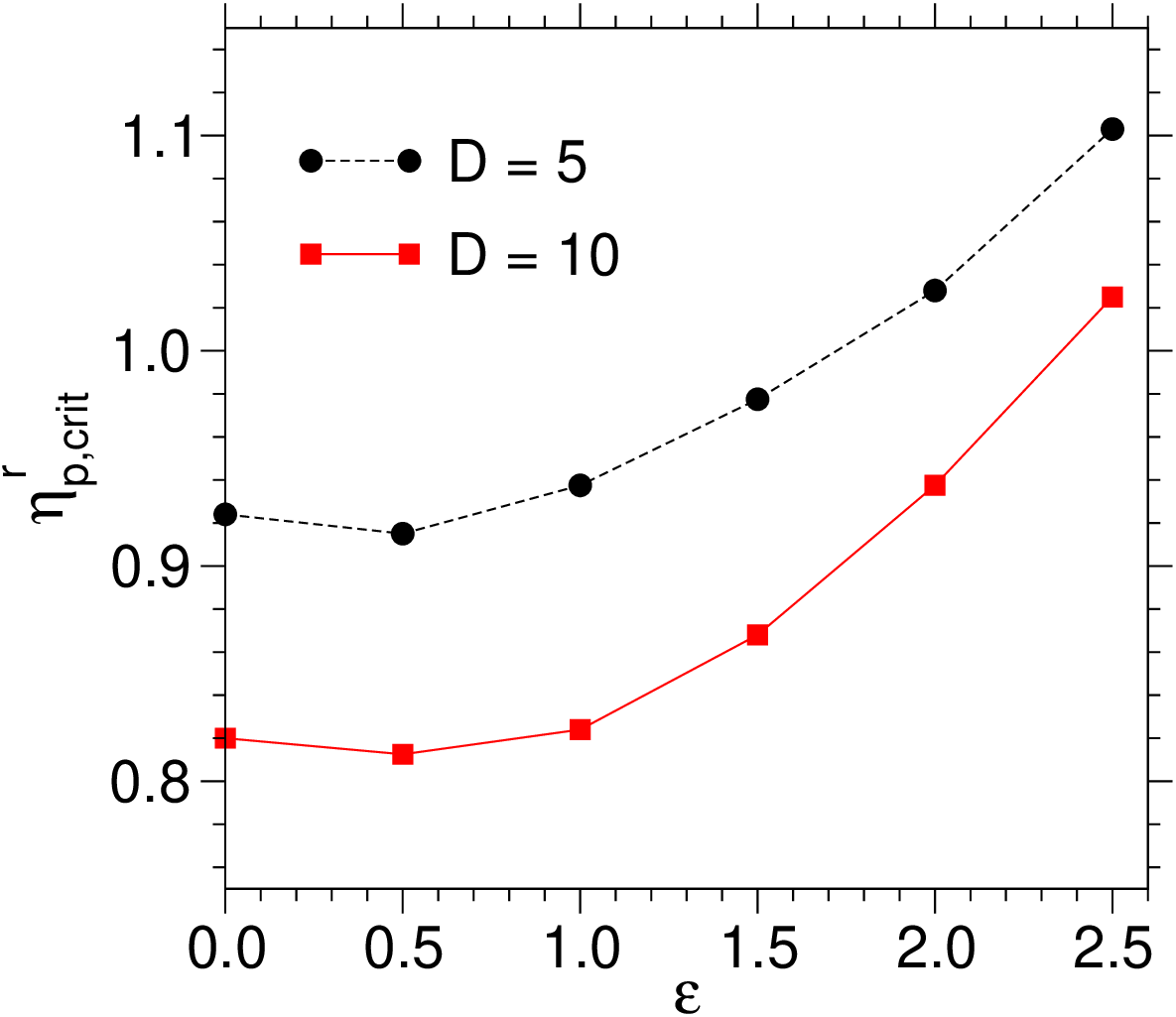}
\caption{Dependence of the critical value $\eta^{\rm r}_{\rm p, crit}$
on $\varepsilon$, for $D=5$ (circles) and $D=10$ (squares). \label{fig5}}
\end{figure}
For $\eta_{\rm p}^{\rm r}=0.95$, on the other hand, the profiles are very
different: phase separation in a colloid--rich and a colloid pure phase
has occurred, separated by an interface in the center of the slit. The
interfacial profile resembles that of an interface between bulk coexisting
phases (broadened by capillary waves \cite{33}). For $\eta^{\rm r}_{\rm
p}=1.2$, finally, the interface is localized either at the right wall
or at the left wall (Fig.~\ref{fig1}c,d): phase separation has occurred
in the thin film (Fig.~\ref{fig2}).

Since Fig.~\ref{fig2} refers to finite $L$, the data for the coexistence
curves do not merge at the critical points, but rather extend somewhat
into the one--phase region (``finite size tails'').  However, repeating
the simulation for several choices of $L$ and performing a finite size
scaling \cite{30} analysis, as done for this model both in the bulk
\cite{28,29} and in thin films with symmetric walls \cite{18,19}, the
critical point can be determined rather reliably. Figure~\ref{fig3}
gives an example for $\varepsilon=1.0$. Due to crossover effects the
cumulant intersections do not occur precisely in a point, but rather
are spread out over some region, but nevertheless, the critical point
$\eta^{\rm r}_{\rm p, crit}$ can be inferred with reasonable accuracy.
The effective exponents $\nu_{\rm eff} \approx 0.947$ (extracted from
$Y_1$, Fig.~\ref{fig2}(a)) and $(\beta/\nu)_{\rm eff} \approx 0.155$
(extracted from $\Delta=(\eta^\ell_{\rm c}-\eta_{\rm c}^v)/2$ at
$\eta^{\rm r}_{\rm p, crit}$ \{not shown\}) deviate somewhat from the
$2d$ Ising values ($\nu =1$, $\beta/\nu=0.125)$, similarly as in thin
Ising films with competing walls \cite{7}.

When $\varepsilon$ increases, $\eta^{\rm r}_{\rm p, crit}$ is shifted to
considerably larger values and at the same time the chemical potential
$\mu_{\rm coex}(D, \varepsilon)$ moves towards the bulk curve,
$\mu^{\rm bulk}_{\rm coex}$ (Fig.~\ref{fig4}), and is almost reached
for $\varepsilon=2.5$. The limiting value $\eta^{\rm r}_{\rm p, crit}
(D \rightarrow \infty)$ for which $\mu_{\rm coex} (D \rightarrow \infty,
\varepsilon)= \mu^{\rm bulk}_{\rm coex}$ would yield an estimate for
the wetting transition $\eta^{\rm r}_{\rm p, w}$ \cite{4,5,7,8,13} of
this model. Very roughly, we estimate $\eta^{\rm r}_{\rm p, w} \approx
1.20 \pm 0.05$.

Comparing to the case of symmetric hard walls $(\varepsilon=0)$, a
non--monotonic variation of $\eta^{\rm r}_{\rm p, crit}$ on $\varepsilon$
is found (Figs.~\ref{fig4}, \ref{fig5}). A similar behavior has been
seen for asymmetric polymer mixtures studied by self--consistent field
theory \cite{11}. Our finding implies that this non--monotonic variation
is a rather general phenomenon, and it holds beyond mean field theory.

\section{Conclusions}
In a colloid--polymer mixture confined by hard walls, depletion attraction
leads to the formation of colloid--rich layers near the walls. By adding
a suitable short range interaction at one of the walls, polymers can be
effectively attracted to this wall whereas the depletion attraction of
colloids is reduced at it. In this case of a colloid--polymer mixture
confined between asymmetric walls, our simulations demonstrate that an
interface localization transition can occur.  Our results also imply that
semi--infinite colloid--polymer mixtures, confined by a hard wall, should
exhibit a wetting transition at sufficiently large polymer fugacity.
We suggest that interface localization transitions could be realized
experimentally by suitable coating of one wall of a slit pore with a
polymer brush.

\acknowledgments
We are grateful to the Deutsche Forschungsgemeinschaft (DFG) for financial
support, project N$^o$'s TR6/A5, D3.

\end{document}